\documentclass[11pt]{article}
\usepackage[left=1in,top=1in,right=1in,bottom=1in,head=0in,nofoot]{geometry}

\usepackage{setspace,url,bm,amsmath} 
\usepackage{scalerel}
\usepackage{xcolor}
 \usepackage{multirow}
 \usepackage{arydshln}
\usepackage{titlesec} 
\titlelabel{\thetitle.\quad} 
\titleformat*{\section}{\bf\Large\center}

\usepackage{eufrak}
\usepackage{float}
\usepackage{graphicx} 
\usepackage{bbm}
\usepackage{latexsym}
\usepackage{caption}
\usepackage[margin=20pt]{subcaption}
\usepackage{enumerate}

\def\ewb{(\e_\bz,\eb)}
\def\ewa{(\e_\az,\ea)}

\usepackage[normalem]{ulem}

\def\T{{ \mathrm{\scriptscriptstyle T} }}
 
\def\fpp{{ F_{\ps[,-1]}}}
\def\fppt{{  F_{\ps[,-1]}^\T}}

\def\zpki{Z'_{i,\mks}}

\def\tkp{\tau_{\mks, \pi} }

\def\dk{\delta_k}

\usepackage{scalerel}
\def\prh{\pr}
\def\by{\bar Y}
\def\hy{\hat Y}
\def\hv{\hat V}
\def\begini{\begin{itemize}}
\def\endi{\end{itemize}}

\def\reg{regression }
\def\regs{regressions }

\def\hchtp{\hat\cov(\hbt_\pi)}
\def\hchtd{\hat\cov(\hbt_{\dprs})}

\def\tsp{\tilde\Psi_{\ps}}

\def\dprs{{\delta\prods}}

\def\dpr{\delta_\prods}

\def\htd{\htau_{\delta\prods}}

\def\kik{{\scaleto{k \in \mk}{6pt}}}

\def\hgp{\hg_\ps}
\def\tgp{\tg_{\ps}}
\def\hgdp{\hg_{\ps}}
\def\hgdm{\hg_{\ms}}
\def\hgm{\hg_\ms}

\def\fa{A}
\def\fb{B}
\def\fc{C}

\def\htwp{ \hbt_{\delta\prods,\ps}}
\def\tpp{\tau_{\pi,\ps}}

\def\twp{ \tau_{\delta\prods,\ps}}
\def\twm{ \tau_{\delta\prods,\ms}}

\def\mfk{\mf_\mks}
\def\mm{m}

\newcommand{\prett}{Under the $2^2$ experiment, the outputs of \eqref{ols_22} 
 satisfy} 
\newcommand{\pretk}{Under the $2^K$ experiment, the outputs of \eqref{ols_2k_g} satisfy}
\newcommand{\pretth}{Under the $2^3$ experiment, the outputs of \eqref{ols_23_g} satisfy} 

\def\hgd{\hat\gamma}

\def\hsd{\hat\Psi}
\def\tgd{\tg_\ps}

\newcommand{\prods}{{\scaleto{\times}{4pt}}}

\newcommand{\tpi}{\tau_\pi}
\newcommand{\fpt}{\fp^\T}

\newcommand{\mk}{ { {\mathcal{K}} }}
\newcommand{\mkc}{{\overline \mk}}

\newcommand{\zk}{{z_\mks}}

\newcommand{\mps}{\mathcal P_{K}}

\def\mks{ {\scaleto{\mathcal{K}}{4.5pt}}}
\def\mkcs{{\kern 0.05em\overline{\scaleto{\mk}{4.5pt}}}}
\def\mkcss{{\scaleto{\overline{\mk}}{5pt}}}
\def\zkc{z_\mkcs}
\newcommand{\zkcs}{ {z_\mkcss }}

\newcommand{\zotokc}{ z_{\mpotk}}

\newcommand{\tokzkc}{\tau_{\mks}(\zkc)}

\DeclareMathOperator{\diag}{diag}
\newcommand{\hY}{\hat Y}

\newcommand{\bt}{\tau}

\newcommand{\htau}{\hat \tau}
\newcommand{\hbt}{\hat \tau}

\newcommand{\tg}{\tilde\gamma}

\newcommand{\qg}{Q_1 \times\cdots \times Q_K}

\newcommand{\bz}{{\B=0}}
\newcommand{\bo}{{\B=1}}
\newcommand{\az}{{\A=0}}
\newcommand{\ao}{{\A=1}}
\newcommand{\cz}{{\C=0}}
\newcommand{\co}{{\C=1}}

\newcommand{\piaz}{\pi_{\A=0}}
\newcommand{\pibz}{\pi_{\bz}}

\newcommand{\piao}{\pi_{\A=1}}
\newcommand{\pibo}{\pi_{\bo}}
\newcommand{\pico}{\pi_{\C=1}}

\newcommand{\fm}{ F_\ms}
\newcommand{\fp}{ F_\ps}

\newcommand{\mf}{{\mathcal{F}}}
\newcommand{\mfp}{{\mf_\ps}}

\newcommand{\ps}{{\scaleto{+}{4pt}}}
\newcommand{\ms}{{\scaleto{-}{4pt}}}

\newcommand{\piab}{\pi_\ab}
\newcommand{\piac}{\pi_\ac}
\newcommand{\pibc}{\pi_\bc}

\newcommand{\e}{e}

\newcommand{\ea}{\e_\ao}
\newcommand{\eb}{\e_\bo}

\newcommand{\da}{\delta_\A}
\newcommand{\db}{\delta_\B}
\newcommand{\dc}{\delta_\C}

\newcommand{\hbb}{\hat{ \beta}}

\newcommand{\ab}{{\A\B}}
\newcommand{\ac}{{\A\C}}

\newcommand{\abc}{{\A\B\C}}
\newcommand{\bc}{{\B\C}}

\newcommand{\pab}{\pi_\ab}
\newcommand{\pac}{\pi_\ac}
\newcommand{\pbc}{\pi_\bc}

\newcommand{\pabc}{\pi_{ab}}
\newcommand{\pacb}{\pi_{ac}}
\newcommand{\pbca}{\pi_{bc}}

\newcommand{\pia}{\pi_\A}
\newcommand{\pib}{\pi_\B}
\newcommand{\pic}{\pi_\C}

\newcommand{\C}{{\textsc{c}}}

\newcommand{\fmh}{\Phi}

\newcommand{\abo}{{\A|\bo}}
\newcommand{\abz}{{\A|\bz}}
\newcommand{\bao}{{\B|\ao}}
\newcommand{\baz}{{\B|\A=0}}
\newcommand{\baa}{{\B|\aaa}}
\newcommand{\abb}{{\A|\bbb}}

\newcommand{\sumz}{\sum_{ z \in \mT}}

\def\otmmo{ {\scaleto{[\mm-1]}{7pt}}}
\def\otm{ {\scaleto{[m]}{7pt}}}
\newcommand{\ktk}{{\scaleto{\mm:K}{5pt}}}
\newcommand{\otk}{{\scaleto{[K]}{6pt}}}
\newcommand{\otkmo}{{\scaleto{[K-1]}{6pt}}}
\newcommand{\mpotk}{{\scaleto{(\mm+1):K}{7pt}}}

\newcommand{\ttk}{{\scaleto{2:K}{5pt}}}

\newcommand{\tabc}{\tau_{\AB}}

\newcommand{\taw}{\tauA(\pib)}
\newcommand{\tbw}{\tauB(\pia)}

\newcommand{\aaa}{{\A = a}}
\newcommand{\bbb}{{\B = b}}
\newcommand{\ccc}{{\C = c}}

\newcommand{\acbz}{{\A\C|\bz}}

\newcommand{\bcao}{{\B\C|\A=1}}
\newcommand{\bcaz}{{\B\C|\A=0}}

\newcommand{\sumi}{\sum_{i=1}^N}
\newcommand{\bd}{ \delta}

\newcommand{\op}{o_\text{p}(1)}

\newcommand{\A}{\textsc{a}}

\newcommand{\B}{\textsc{b}}

\newcommand{\AB}{\textsc{ab}}

\newcommand{\tauA}{\tau_\textsc{a}}
\newcommand{\tauB}{\tau_\textsc{b}}
\newcommand{\tauAB}{\tau_\textsc{ab}}

\newcommand{\pr}{{\rm pr}}

 \newcommand{\ot}[1]{1, \ldots,#1}

\def\T{{ \mathrm{\scriptscriptstyle T} }}

\newcommand{\cov}{\textup{cov}}

\newcommand{\hg}{\hat\gamma}

\newcommand{\bY}{{\bar Y}}

\newcommand{\hV}{\hat V}
\newcommand{\hbV}{\hat{V}}

\newcommand{\mT}{\mathcal{T}}
\newcommand{\mt}{\mathcal{T}}

\def\begina{\begin{eqnarray*}}
\def\enda{\end{eqnarray*}}
\def\beginy{\begin{eqnarray}}
\def\endy{\end{eqnarray}}
\def\begine{\begin{enumerate}}
\def\ende{\end{enumerate}}

\usepackage[utf8]{inputenc}
\usepackage[english]{babel}

\usepackage{amsthm}
\usepackage{amssymb}
\usepackage{amsmath}
\usepackage{color}

\usepackage{comment}
\theoremstyle{definition}

\newtheorem*{theorem*}{Theorem}
\newtheorem{theorem}{Theorem}
\newtheorem*{rmk*}{remark}
\newtheorem{proposition}{Proposition}

\newtheorem{condition}{Condition}

\newtheorem{definition}{Definition}
\newtheorem{remark}{Remark}
\newtheorem{corollary}{Corollary}
\newtheorem*{corollary*}{Corollary}

\usepackage{natbib} 
\bibpunct{(}{)}{;}{a}{}{,} 

\usepackage{etoolbox} 
\apptocmd{\sloppy}{\hbadness 10000\relax}{}{} 

\usepackage{multibib}
\newcites{sec}{References}

\usepackage{color}
\usepackage{listings}
\usepackage{hyperref}
\usepackage{booktabs}
\usepackage{lscape}

\RequirePackage[normalem]{ulem}

\allowdisplaybreaks

\begin{document}
\onehalfspacing

\title{\bf \Large
Regression-based causal inference  with factorial experiments: estimands, model specifications, and design-based properties}
\author{Anqi Zhao and Peng Ding
\footnote{Anqi Zhao, Department of Statistics and Data Science, National University of Singapore, 117546, Singapore (E-mail: staza@nus.edu.sg). Peng Ding, Department of Statistics, University of California, Berkeley, CA 94720 (E-mail: pengdingpku@berkeley.edu).
}
}
\date{}

\maketitle

\begin{abstract}
Factorial designs are widely used due to their ability to accommodate  multiple factors simultaneously.  
Factor-based regression with main effects and some interactions is the dominant strategy for downstream analysis, delivering point estimators and standard errors simultaneously via one least-squares fit. 
Justification of these convenient estimators from the design-based perspective requires quantifying their sampling properties under the assignment mechanism whilst conditioning on the potential outcomes.  
To this end, we derive the sampling properties of the regression estimators under a wide range of specifications, and establish the appropriateness of the corresponding robust standard errors for the Wald-type inference. 
The result clarifies the causal interpretation of the coefficients in these factor-based regressions, and inspires the definition of general factorial effects to unify the standard definitions of factorial effects in various literatures.
We also quantify the  bias-variance trade-off between the saturated and unsaturated regressions 
from the design-based perspective. 
\end{abstract} 


\section{Introduction}

Factorial designs are increasingly popular in field experiments in social sciences \citep[e.g.,][]{duflo2007using, td, Branson2016, imai} in addition to the traditional agricultural, industrial, and biomedical applications \citep[e.g.,][]{wh}. 
Factor-based regression remains the dominant strategy for downstream analysis \citep[e.g.,][]{Karlan2007, Eriksson2014, torres}, 
enabling not only direct estimation of the factorial effects as regression coefficients
but also flexible unsaturated specifications to reduce model complexity.  A formal justification of its role in causal inference, however, requires both clearly defining the estimands of interest and deriving the sampling properties of the resulting estimators under the potential outcomes framework. 

This article makes several contributions.
First, we clarify the causal interpretation of the coefficients in factor-based linear regressions and
propose a location-shift strategy to reproduce the design-based inference of various factorial effects via least squares.
Importantly, we show that the robust covariance affords an asymptotically conservative estimator of the true sampling covariance from the design-based perspective, justifying its use for large-sample Wald-type inference. 
Second, we review and clarify the standard definitions of factorial effects in the causal inference, experimental design, epidemiology, and social sciences literatures, and extend them to allow for arbitrary weighting schemes to accommodate external validity concerns. 
Third, we derive the design-based properties of estimators from unsaturated factor-based regressions for the first time, and quantify the bias-variance trade-off between the saturated and unsaturated regressions from the design-based perspective.

We use $Y_i \sim x_i$ to denote the least-squares regression of $Y_i$ on $x_i$ and focus on not only the causal interpretation of the regression coefficients for estimating the general factorial effects but also the design-based properties of the robust covariance, also known as the Eicker--Huber--White covariance,  for large-sample Wald-type inference. 
The terms ``regression", ``coefficients", and ``robust covariance" refer to the numeric outputs of least squares free of any modeling assumptions; we evaluate their sampling properties from the design-based perspective. 
We omit the discussion of the ordinary covariance derived under homoskedasticity  
due to its lack of 
design-based guarantees even with the simple treatment-control experiment \citep{Freedman08a}.

Let $1_N$ and $1_Q$ be the $N\times 1$ and $Q\times 1$ vectors of ones, respectively.
Let $ \mathcal{I}(\cdot)$ be the indicator function.  
Let $[m] = \{\ot{m}\}$ be the set of $1$ to $m$ for positive integer $m$. 
For two symmetric matrices $M_1$ and $M_2$, write $M_1 \geq 0$ if $M_1$ is positive semi-definite and write $M_1 \leq M_2$ or $M_1 \geq M_2$ if $(M_2-M_1)$ is positive or negative semi-definite, respectively.

%
\section{Framework, causal effects, and treatment-based regression}\label{sec:treatment-based}

Consider an experiment with $N$ units, $i = 1, \ldots, N$, and  $Q$ treatment levels, $ z \in \mt = \{\ot{Q}\}$. 
Let $Y_i(z)$ be the potential outcome of unit $i$ if assigned to level $z  $, and let $\bar Y(z) = N^{-1} \sumi  Y_i(z)$ be the average, vectorized  as $\bY = (\bY(1), \ldots, \bY(Q) )^\T$. Let $ S =  (S(z,z') )_{z,z'\in\mt}$ be the finite-population covariance matrix of the potential outcomes  with
$S(z,z') = (N-1)^{-1}\sumi \{Y_i(z)-\bY(z)\}\{Y_i(z')-\bY(z')\}$. 
The goal is to estimate $
\tau = G \bY$
for some contrast matrix $G$  with rows orthogonal to $1_Q$. 
Complete randomization assigns completely at random $ N_z \geq 2$  
units to level $z $ with $\sum_{z\in\mt} N_z = N$ and $e_z = N_z/N$.  
For unit $i$, let $Z_i \in \mt$ denote the treatment level and $Y_i = \sumz \mathcal{I}(Z_i = z) Y_i(z)$ denote the observed outcome.
Let $\hat Y(z) = N_z^{-1}\sum_{i: Z_i = z} Y_i$  be the average observed outcome under level $z$, vectorized as $\hY = (\hY(1), \ldots, \hY(Q) )^\T$.  Then $\hbt=G\hY$ affords an intuitive choice for estimating $\tau$.

Design-based inference, also known as the randomization inference, concerns the sampling properties of estimators  over
the distribution of the treatment indicators,  
conditioning on the potential outcomes. Throughout the paper, we focus on complete randomization and invoke Condition \ref{asym} below for asymptotic properties \citep{DingCLT}.

\begin{condition}\label{asym}
As $N$ goes to infinity, for all $z \in \mt$, (i) $ N_z \geq 2$  
and $e_z$ has a limit between $(0,1)$, (ii) $\bY$ and $ S$ have finite limits, and (iii) $\max_{1\leq i \leq N} \{Y_i(z) - \bar Y(z)\}^2/N  \rightarrow 0$. 
\end{condition}

Under complete randomization, $\hY$ is unbiased for $\bY$ with covariance 
$\cov( \hY ) = \diag \{ S(z,z)/N_z \}_{z\in\mt}  - N^{-1} S$. 
Define $\hbV = \diag \{\hat S(z,z)/N_z \}_{z\in\mt}$, where $\hat S(z,z) = (N_z-1)^{-1}\sum_{i: Z_i = z} \{Y_i - \hY(z)\}^2$, as a moment estimator of $\cov(\hY)$. 
It is conservative in the sense of $ E(\hV) - \cov( \hY )  = N^{-1} S \geq  0$.
Condition \ref{asym} further ensures  $\hY$ is asymptotically Normal with $N\{ \hV - \cov( \hY ) \} =  S + \op$ \citep{DingCLT}. 
The Wald-type inference of $\tau$ can thus be conducted using $\hbt = G\hY$ and   $\hat\cov(\htau) = G\hV G^\T$ as the point estimator and estimated covariance, respectively. 
It is in general conservative due to the over-estimation of the covariance; one exception is when the   treatment effects are constant across all units as specified by Condition \ref{cond:sa} below.

\begin{condition}
\label{cond:sa} 
For all $z,z'\in\mt$, $Y_i(z) - Y_i(z') = c(z,z')$
are
constant across $i = \ot{N}$. 
This ensures $S(z,z')$ are identical for all $z,z'\in\mt$, denoted by $S(z,z') = s_0$.
\end{condition}

Treatment-based regression  affords a convenient tool for computing $\hY$ and $\hat V$ from least squares. 
The regression 
$
Y_i \sim   \mathcal{I}(Z_i =  1) + \cdots + \mathcal{I}(Z_i =  Q) 
$
without an intercept yields coefficient vector $\hbb$ and robust covariance $\hV_0 $ that satisfy $\hbb = \hY$ and $\hV_0 = \diag(1-N_z^{-1})_{z\in\mt}  \hV = \hV + \op $ \citep[][Section 3.3]{D20}.  
The Wald-type inference of $\tau$ can thus also be conducted using $G\hat\beta$ and $G\hat V_0 G^\T$ as the point estimator and estimated covariance, respectively.

As a special case, this setup encompasses the $\qg$ factorial experiment, which
involves $Q = \prod_{k=1}^K Q_k$ treatment levels as the combinations of $K\geq 2$ factors  with $Q_k \ (k = \ot{K})$ levels, respectively. Treatment-based regression accordingly 
affords a principled way to study general factorial experiments. It is nevertheless  not the dominant strategy in practice when the estimands of interest take some special forms. In the case where the goal is to estimate the main effects or interactions of the factors under study, a more prevalent practice is to regress the outcome on the factors themselves and interpret the coefficients as the corresponding factorial effects of interest. This seemingly straightforward approach has several variants across different fields, which turn out to target factorial effects under distinct weighting schemes. 
Our first contribution unifies these variants under a class of location-shifted factor-based regressions, and establishes the design-based properties of  the resulting coefficients and robust covariances.

More importantly,  treatment-based regression is saturated and requires the estimation of $  Q = \prod_{k=1}^K Q_k \geq 2^K$ parameters. 
This could be demanding in terms of sample size even with a moderate number of factors.
Factor-based regression, on the other hand, enables flexible unsaturated specifications that include only the main effects and possibly some lower-order interactions corresponding to the factorial effects of interest. 
Despite the intuitiveness of such an approach and its dominance in practice, 
the existing literature on the design-based properties of factor-based regression focuses on saturated specifications \citep{td, FactLu16b}, and leaves the theory of their unsaturated counterparts 
an open question.   
Our second contribution fills this gap  and establishes the design-based properties of unsaturated factor-based regressions.

Due to the notational burden involved in the general setting, we start with the $2^2$ and $2^3$ experiments to illustrate the main ideas, and then  unify the  results under the $2^K$ experiment. The results convey all key points for the theory of the general $\qg$ experiment.  We give the formal theory on the general case in the Supplementary Material.

\section{The $2^2$ factorial experiment}\label{sec:22}
%
\subsection{A review of existing strategies}\label{sec::review-strategies-22}
The $2^2$ factorial experiment is the simplest factorial experiment with two binary factors,  A  and  B. 
The $Q=2^2 = 4$ treatment combinations consist of $\mt = \{(00), (01),(10),  (11) \}$, indexed by $z = (ab)$ for $a,b=0,1$. 
 Let $A_i, B_i \in \{0,1\}$ indicate the levels of the factors received by unit $i$. 
We first review five factor-based regression strategies commonly used for  analyzing $2^2$ experiments, and then clarify their respective causal interpretations. 

The canonical  factor-based regression takes the form  $Y_i \sim 1 + A_i + B_i + A_i  B_i$. 
Strategy (i) directly uses the coefficients of $(A_i, B_i, A_iB_i)$, denoted by $\hg_0 = (\hg_{0,\A}, \hg_{0,\B}, \hg_{0,\AB})^\T$, 
to estimate the main effects of factors {\fa} and {\fb} and their interaction, respectively. 
Strategy (ii) uses $( \hg_{0,\A} +  B_i\hg_{0,\AB},   \hg_{0,\B} +  A_i\hg_{0,\AB},  \hg_{0,\AB})$   
to estimate the main effects and interaction at the unit level, respectively,
and then takes their respective averages  
to estimate the factorial effects at the population level. Define $\ea  = N^{-1}\sumi A_i$ and $\eb = N^{-1}\sumi B_i$ as the empirical probabilities of  factors A and B, respectively.
The final estimators equal $\hg_{\textup{e}} = (\hg_{\textup{e},\A}, \hg_{\textup{e},\B}, \hg_{\textup{e},\AB})^\T$ where $\hg_{\textup{e},\A} = \hg_{0,\A} +  \eb\hg_{0,\AB}$, $\hg_{\textup{e},\B} = \hg_{0,\B}+  \ea\hg_{0,\AB}$, and $\hg_{\textup{e},\AB} = \hg_{0,\AB}$.
Strategy (ii) is popular in econometrics, with the estimators of the main effects, namely $\hg_{\textup{e},\A}$ and $  \hg_{\textup{e},\B} $, also known as the average partial or marginal effects  \citep{greene}.
Strategy (iii) codes the factors by their signs as $   A_i^{\scaleto{\textup{s}}{4pt}} = 2A_i -1$ and $  B_i^{\scaleto{\textup{s}}{4pt}}= 2B_i -1 \in \{+1, -1\}$, 
and uses the coefficients from $
Y_i \sim 1+ A_i^{\scaleto{\textup{s}}{4pt}} +B_i^{\scaleto{\textup{s}}{4pt}}  +A_i^{\scaleto{\textup{s}}{4pt}} B_i^{\scaleto{\textup{s}}{4pt}}$, after
multiplied by two, to estimate the main effects and interaction, respectively \citep{wh, FactLu16b}. Let $\hg_{\textup{s} }  =   (\hg_{\textup{s}, \A }, \hg_{\textup{s}, \B }, \hg_{\textup{s}, \AB })^\T$ denote the estimators under strategy (iii). This gives three strategies for simultaneously estimating the main effects and interaction via one least-squares fit.

Strategies (iv) and (v), on the other hand, focus on only the two main effects. 
Strategy (iv) considers two separate regressions, $Y_i \sim 1+ A_i$ and $Y_i \sim 1+ B_i$, and estimates the two main effects by the coefficients of $A_i$ and $B_i$, respectively \citep[e.g.,][]{bertrand2004emily, Eriksson2014}. 
Strategy (v) considers the additive regression $Y_i \sim 1+A_i + B_i$, and estimates the two effects via one least-squares fit.

Refer to a factor-based regression as saturated if it contains all possible interactions between the factors in addition to the constant term and main effects.
The regressions under strategies (i)--(iii) are saturated whereas those under strategies (iv) and (v) are unsaturated.

\subsection{Unifying the saturated regressions and introducing the general factorial effects}\label{sec::unification22}
We now unify strategies (i)--(iii) under a class of location-shifted  
factor-based regressions that turn out to target  factorial effects under different weighting schemes. 
The result highlights the correspondence between the location shifts in specifying the models and the weighting schemes in defining the factorial effects. 

To this end, we first formalize the notion of general factorial effects, which are central to clarifying the effective estimands under strategies (i)--(iii).
Define $\tau_{\A | b} =\tau_\abb =  \bar Y(1b) - \bar Y(0b)$ and $\tau_{\B |a} =\tau_\baa = \bar  Y(a1) -  \bar Y(a0)$ as the conditional effects of factors {\fa} and {\fb} when the level of the other factor is fixed at $b \in \{0,1\}$ and $a\in \{0,1\}$, respectively.
As a convention, we abbreviate the ``$\aaa$" and ``$\bbb$" in the subscripts as ``$a$" and ``$b$", respectively, when no confusion would arise. 
Define
\begina
\taw  = \pibz \cdot \tau_\abz + \pibo \cdot \tau_\abo,\qquad 
\tbw  =  \piaz \cdot \tau_\baz +  \piao\cdot \tau_\bao
\enda
as the main effects of factors {\fa} and {\fb} under weighting schemes  $\pib=(\pibz,\pibo)$ and $\pia = (\piaz,\piao)$, respectively, with 
$0 \leq \pi_\aaa, \pi_\bbb \leq 1$   for $a,b=0,1$  and $  \piaz+\piao= \pibz+\pibo=1$.
As a convention, the subscript of the weighting scheme indicates the factor that is being marginalized out.
The standard main effects
correspond to $\pia = \pib = (1/2, 1/2)$, weighting all conditional effects equally \citep{td}.

Define $\tabc = \bY(11)-\bY(10)  -  \bY(01) + \bY(00)$  as the interaction between {\fa} and {\fb}.
It satisfies 
$
\tabc 
= \tau_{\A|\bo} - \tau_{\A|\bz}
= \tau_{\B|\ao} - \tau_{\B|\az}
$ 
and characterizes the difference in the conditional effects of one factor at the two levels of the other factor.
Note that $\tauA(\pib') - \tauA(\pib) = (\pibo' - \pibo) \tauAB$ and 
$\tauB(\pia') - \tauB(\pia) = (\piao' - \piao) \tauAB$ such that $\tau_\AB$ also quantifies the difference in causal estimands between different weighting schemes. 
The absence of the interaction, namely $\tau_\ab=0$, ensures that 
$\tauA(\pib) = \tau_{\abz}$ and $\tauB(\pia) = \tau_{\baz}$ are constant across all possible weighting schemes.

Recall $\bY= (\bY(00),\bY(01), \bY(10),\bY(11))^\T$. Vectorize the main effects and interaction as  $
\bt_\pi = (\taw, \tbw,  \tauAB)^\T =  G_\pi \bY$ with $\pi =(\pia, \pib) $ and the contrast matrix $G_\pi $ consisting of row vectors  $  (-\pibz, -\pibo, \pibz, \pibo ) $,  $  (-\piaz, \piaz, -\piao, \piao ) $, and $  (1, -1, -1, 1) $. An unbiased estimator for $\bt_\pi$ is $\hbt_\pi = G_\pi \hY = (\htau_\A(\pib), \htau_\B(\pia), \htau_\AB)^\T$.

Let $e_\az = 1-\ea$ and $e_\bz =  1-\eb$ be the proportions of units that receive level 0 of factors A and B in the experiment, respectively.
 Proposition \ref{prop:123} below is numeric and clarifies the causal interpretations of the regression estimators from strategies (i)--(iii). 

\begin{proposition}\label{prop:123}
Under the $2^2$ experiment, the coefficients from strategies \textup{(i)}--\textup{(iii)} satisfy 
\begine 
\item[\textup{(i)}]  $ \hg_{0}  = (\htau_\A(1,0), \htau_\B(1,0), \htau_\AB)^\T$ with $\pi_\A =\pi_\B = (1, 0)$;
\item[\textup{(ii)}] $\hg_{\textup{e}}  =  (\htau_\A(\e_\bz,\eb), \htau_\B(e_\az,\ea), \htau_\AB)^\T$ with 
$\pi_f = (\e_{f=0}, e_{f=1})$ for $f = \A, \B$;  
\item[\textup{(iii)}] $ \hg_{\textup{s} }   =  (\htau_\A(1/2,1/2), \htau_\B(1/2,1/2), \htau_\AB/2)^\T$ with $\pi_\A =\pi_\B = (1/2, 1/2)$.
\ende
\end{proposition} 

Strategies (i)--(iii) thus yield identical estimators of $\tauAB$ up to a scaling factor yet target at distinct main effects under different weighting schemes. 
Strategy (i) is unbiased for estimating $\tauA(1,0) = \tau_\abz$ and $\tauB(1,0) = \tau_\baz$ as the conditional effects when the other factor is at the baseline level. 
Strategy (ii) is unbiased for estimating $\tauA\ewb$ and $\tauB\ewa$; the average partial effects in econometrics thus weight the conditional effects by the empirical treatment probabilities.   
Strategy (iii) is unbiased for estimating the standard effects $\tauA = \tauA(1/2, 1/2)$ and $\tauB= \tauB(1/2, 1/2)$ that weight all conditional effects equally.   
This clarifies the causal interpretations of $\hg_0$, $\hg_{\textup{e}}$, and $\hg_{\textup{s}}$ from strategies (i)--(iii), respectively.
In particular, $\hg_{\textup{s}}$ targets the standard factorial effects regardless of whether the experiment is balanced or not.

Inspired by how transformation on factors allows us to obtain the moment estimators of the standard main effects directly as regression coefficients under strategy (iii), we now propose a location-shift strategy to generalize strategies (i)--(iii) and estimate $\tau_\pi$ with arbitrary weights $\pi = (\pia, \pib)$ via least squares. 
For $ A_i' =A_i-\da$ and $ B_i' =B_i-\db$ with prespecified  $0 \leq \da, \db\leq 1$, 
define the location-shifted regression
\begin{eqnarray}
\label{ols_22}
Y_i \sim 1 + A_i' + B_i' + A_i ' B_i'
\end{eqnarray}
with coefficients $\hgd = (\hg_{\A}, \hg_{ \B}, \hg_{ \ab})^\T$ and robust covariance $\hsd$ for the three non-intercept terms.
Strategies (i)--(iii) are special cases: setting $(\da, \db) = (0,0)$ equals strategy (i); 
setting $(\da, \db) = (\ea, \eb)$ equals strategy (ii) in the sense of $\hgd = \hg_\textup{e}$ by Proposition \ref{prop:22} below;
setting $(\da, \db) = (1/2, 1/2)$ equals strategy (iii)  up to scaling factors of two or four.

Recall $\htau_\pi = G_\pi\hy$ as an unbiased estimator of $\tau_\pi= G_\pi \by$.  
Let $\hchtp  =   G_\pi \hV G_\pi^\T$ be the corresponding estimated covariance, recalling $\hv$ as a conservative estimator of $\cov(\hy)$.  
Proposition \ref{prop:22} below  states the numeric correspondence between $\{ \hg, \hat\Psi \}$ and $\{ \htau_\pi, \hchtp \} $,
elucidating the design-based properties of $\hg$ and $\hsd$ for general $(\da, \db)$.

\begin{proposition}\label{prop:22}
{\prett} 
 $\hgd = \htau_\pi$ and $\hsd  =  \hchtp - G_\pi   \diag(  N_z^{-1})   \hV   G_\pi^\T$
for $\pi =(\pia, \pib)$ with $\pia = (1-\da,\da)$ and $\pib = (1-\db,\db)$. 
\end{proposition}

Proposition \ref{prop:22} ensures that $\hat\gamma$ from \eqref{ols_22} is unbiased for estimating $\tau_\pi$ with $\pia = (1-\da,\da)$ and $\pib = (1-\db,\db)$. 
Location shifts of $A_i$ and $B_i$ by $(\da,\db) = (\piao,\pibo)$ thus enable direct estimation of $\bt_\pi$ from \eqref{ols_22} for arbitrary $\pi$. 
This gives the intuition for requiring $ 0\leq \delta_\A, \delta_\B \leq 1$ introduced before. 
Moreover, the difference between $\hsd $ and $\hchtp$ diminishes as $N$ goes to infinity. 
This enables the large-sample Wald-type inference of $\tau_\pi$ by using $\hgd$ and $\hsd$ as the point estimator and estimated covariance, respectively. 

\begin{remark}\label{rmk:bb}
The classical experimental design literature focuses mostly on the standard main effects \citep{wh}, with equal weights on all conditional effects: $\tau_\A = 2^{-1}(\tau_\abz + \tau_\abo)$ and $\tauB =2^{-1} (\tau_\baz + \tau_\bao)$. The standard main effects, together with balanced experiments with $N_z=N/Q$ for all $z\in \mt$, have many advantages in practice. Corollary \ref{corollary::orthogonality_2k} later states a result for the $2^K$ experiment with a general $K$. 

Applications in practice, however, may not  always value $\tau_\abz$ and $\tau_\abo$, and likewise $\tau_\baz$ and $\tau_\bao$, equally. 
Alternative weighting schemes based on perceived importance could thus also merit attention and afford possibly more relevant summary of the marginal effects \citep{finney1948main}. We give an example based on consideration of external validity of the experimental results. 

Assume the experiment in question is a pilot study for a large-scale implementation that intends   $1/3$ of the population to receive level $1$ of factor {\fb} marginally.
Now that we know $2/3$ of the population will be experiencing the effect of factor {\fa} at the baseline level of factor {\fb}, the general effect $\tauA (2/3, 1/3) = 2/3\cdot \tau_\abz +  1/3\cdot \tau_\abo$ can be a better summary of the effect of factor {\fa} compared with the standard effect with equal weights.  
This illustrates the connection between the general weighting schemes and external validity.

When $\tabc \neq 0$, we are also interested in finding the optimal level of factor B to maximize the effect of factor A. This requires us to compare $\tau_\abo$ and $\tau_\abz$, which correspond to two special estimands $\tauA (0,1 ) $ and $\tauA (1,0 ) $.

In summary, the choice of estimand depends on the scientific question of interest. We provide the theory for the general estimand which includes the above examples as special cases. 

\end{remark}

\subsection{Factor-based regression with unsaturated models}\label{sec:us_22}
Strategies (iv) and (v) concern only the main effects of factors A and B. 
To this end, strategy (iv) fits two separate regressions for estimating the main effects of factors A and B, respectively. 
The resulting estimators equal the differences in means between $\{Y_i: f_i = 1\}$ and $\{Y_i: f_i = 0\}$ for $f=A, B$, respectively, and are biased for estimating factorial effects of the form $\tauA(\pib)$ and $\tauB(\pia)$ in general.
We thus exclude it from the ensuing discussion.

Strategy (v), on the other hand, estimates the two main effects together via one additive regression.  
Consider a generalized version, incorporating the location-shift transformation:
\beginy\label{us_22}
Y_i \sim 1+  A^\prime _i +  B^\prime _i .
\endy
We first derive the effective estimands of \eqref{us_22} as a pair of general factorial effects, and then state the bias-variance trade-off between \eqref{ols_22} and  \eqref{us_22}. The result establishes the optimality of \eqref{us_22} for estimating arbitrary $\tau_\pi$ when the nuisance effect $\tau_\ab$ indeed does not exist.

Let $\tg_\A$ and $\tg_\B$ be the coefficients of $A_i'$ and $B_i'$ from \eqref{us_22}, respectively. 
Let $ \hat\tau_{\A|\bbb}$ and $\hat\tau_{\B |\aaa}$ be the moment estimators of $ \tau_{\A|\bbb}$ and $ \tau_{\B |\aaa} $ for $a, b = 0,1$, respectively. 

\begin{proposition}\label{prop:cf_22}
Under the $2^2$ experiment, the coefficients  from \eqref{us_22} satisfy 
\begina
\tg_\A = \tilde{\pi}_{\bz} \cdot  \hat\tau_{\A| \bz } +   \tilde{\pi}_{\bo} \cdot  \hat\tau_{\A| \bo }, \qquad 
\tg_\B = \tilde{\pi}_{\az} \cdot  \hat\tau_{\B | \az } +   \tilde{\pi}_{\ao} \cdot  \hat\tau_{\B | \ao }
\enda with  
$
 \tilde\pi_\B =  (\tilde{\pi}_{\B =0}  ,  \tilde{\pi}_{\B =1}) 
= \sigma^{-1} ( e_{01}^{-1}+ e_{11}^{-1} ,  e_{00}^{-1}+ e_{10}^{-1})$ and $ 
 \tilde\pi_\A =  ( \tilde{\pi}_{\A= 0},  \tilde{\pi}_{\A =1}) 
=\sigma^{-1} ( e_{10}^{-1}+ e_{11}^{-1}, e_{00}^{-1}+ e_{01}^{-1})$,  
where $\sigma = \sum_{z\in\mt} e_z^{-1}$. 
\end{proposition}

Proposition \ref{prop:cf_22}  shows $ \tg_\A$ and $ \tg_\B$ as the moment estimators of $\tau_\A( \tilde\pi_\B)$ and $\tau_\B( \tilde\pi_\A)$ under a specific weighting scheme 
that is fully determined by $(e_{z})_{z\in\mt}$ and independent of $(\da, \db)$.  
Therefore, the unsaturated regression \eqref{us_22} no longer accommodates flexible weighting schemes even with location-shifted factors. 
Under balanced designs with equal treatment sizes $N_z = N/4$ for all $z \in\mt$, we have $\tilde \gamma_\A = \htau_\A(1/2, 1/2)$ and  $\tilde \gamma_\B = \htau_\B(1/2, 1/2)$ give the moment estimators of the standard main effects, and thus equal 
the coefficients of $A_i'$ and $B_i'$ from the saturated regression \eqref{ols_22} with $\delta_\A = \delta_\B = 1/2$.
This is no coincidence but due to the fact that the columns of the design matrix of \eqref{ols_22} with $\delta_\A = \db = 1/2$ are mutually orthogonal such that the deletion of $A_i' B_i'$ has no effect on the estimation of the remaining coefficients. 
This highlights the connection between standard effects and balanced designs from a different angle, echoing the  classical principle  that recommends the use of balanced designs whenever possible.

In general, $ \tg_\A$ and $ \tg_\B$ are biased for $\tau_\A(\pi_\B)$ and $\tau_\B(\pi_\A)$ unless $(\pi_\A, \pi_\B) = (\tilde\pi_\A,  \tilde\pi_\B)$ or the interaction $\tau_\ab$ does not exist. 
Nevertheless, under Condition \ref{cond:sa},  they minimize the sampling variances 
of $\htau_\A(\pi_\B) = {\pi}_\bz  \cdot  \hat\tau_{\A| \bz } +  {\pi}_\bo   \cdot  \hat\tau_{\A| \bo }$ and $\htau_\B(\pi_\A) = {\pi}_\az    \cdot  \hat\tau_{\B | \az } +  {\pi}_\ao   \cdot  \hat\tau_{\B | \ao }$ over all possible $\pi_\B$ and $\pi_\A$, respectively.
In particular, the constant treatment effects ensure $ \textup{var}(\hat\tau_{\A| \bz }) = s_0 (N_{00}^{-1} + N_{10}^{-1})$ and $ \textup{var}(\hat\tau_{\A| \bo }) = s_0 (N_{01}^{-1} + N_{11}^{-1})$.
To minimize the variance of $\htau_\A(\pi_\B)$ is thus equivalent to having the weights proportional to the inverses of $\textup{var}(\hat\tau_{\A| \bz })$ and $ \textup{var}(\hat\tau_{\A| \bo })$, respectively, resulting in $(\tilde{\pi}_\bz   ,\tilde{\pi}_\bo  )$ as defined in Proposition \ref{prop:cf_22}. 
Similar discussion extends to $\tg_\B$. 
This demonstrates  the bias-variance trade-off between \eqref{ols_22} and \eqref{us_22}.

This concludes our discussion on the $2^2$ experiment.
We next extend the results to the $2^3$ experiment to illustrate one additional point: 
 with more than two factors, the factor-based regression is capable of estimating 
 only a subset of all causally-meaningful factorial effects in general, yet regains generality in the absence of three-way interactions.

%
\section{The $2^3$ factorial experiment}\label{sec::23experiment}
%

\subsection{Notation and definition of the general factorial effects}
The $2^3$ factorial experiment features $Q = 2^3 = 8$ treatment combinations arising from three binary factors, A, B, and C.   Let $A_i$, $B_i$, and $C_i \in \{0,1\}$ indicate the levels of the factors  received by unit $i$. 
The eight treatment combinations consist of $ \mt = \{(abc): a,b,c = 0,1\}$.
Let $\by(abc)$ be the average potential outcome under treatment combination $(abc)\in \mt$. 
Define the conditional  effects of factors {\fa}, {\fb}, and {\fc} as
$$
\tau_{\A| bc}  = \bar Y(1bc) - \bar Y(0bc),\quad 
\tau_{\B|ac}  =     \bar Y(a1c) - \bar Y(a0c),\quad 
\tau_{\C|ab}  =  \bar Y(ab1) - \bar Y(ab0) , 
$$ 
respectively, with the rest two factors fixed at $bc, ac, ab \in \{0,1\}^2$. Define the conditional two-way interactions between factors {\fa} and {\fb}, factors {\fa} and {\fc}, and  factors {\fb} and {\fc} as
\begina
\tau_{\ab|c} &=&   \bar Y(11c) - \bar Y(10c) - \bar Y(01c)+\bar Y(00c),\nonumber\\
\tau_{\ac|b}  
&=&  \bar Y(1b1) - \bar Y(1b0)- \bar Y(0b1) + \bar Y(0b0),\label{eq:ce_23}\\
\tau_{\bc|a}  
&=&  \bar Y(a11) - \bar Y(a10)- \bar Y(a01) + \bar Y(a00)\nonumber, 
\enda
respectively, with the third factor  fixed at $c, b, a \in \{0,1\}$. 
When potential confusion arises, we write out ``$\aaa$",  ``$\bbb$", and ``$\ccc$" for $a$, $b$, and $c$ in the subscripts to emphasize both the factors and their respective levels; for example, $\tau_{\A| bc}  = \tau_{\A| \bbb,  \ccc}$ and $\tau_{\ab|c}  = \tau_{\ab|\ccc} $.
These conditional effects afford the building blocks for defining the general factorial effects. 

To simplify the presentation, we call a set of $W$ numbers, $(\pi_1, \ldots, \pi_W)$, a $W$-dimensional  weighting vector  if $\sum_{w=1}^W \pi_w=1$ and $\pi_w \geq 0$; 
a  weighting scheme  is then a collection of weighting vectors with composition clear from the context. 
Throughout this section, assume 
$
\pab = (\pi_{ab})_{ a,b=0,1}, 
\pac = (\pi_{ac})_{a,c=0,1}, 
\pbc = (\pi_{bc})_{b,c=0,1}
$
are some prespecified four-dimensional weighting vectors, and 
$
\pia=(\pi_a)_{a=0,1}, 
\pib=(\pi_b)_{b=0,1}, 
\pic=(\pi_c)_{c=0,1}
$
are some prespecified two-dimensional weighting vectors.
Summarize them as 
$
\pi = \{ \pab, \pac, \pbc, \pia, \pib, \pic\} = \{\pi_{ab}, \pi_{ac}, \pi_{bc}, \pi_a, \pi_b, \pi_c: a, b, c = 0,1\}$.  

\begin{definition}\label{def_23}
Under the $2^3$ experiment, define
$$
   \tau_\A( \pbc) = \sum_{b,c} \pbca \cdot \tau_{\A| bc}, \quad 
 \tau_{\B}(  \pac) =  \sum_{a,c} \pacb \cdot \tau_{\B|ac}, \quad
   \tau_{\C}(  \pab) = \sum_{a,b} \pabc \cdot \tau_{\C| ab}
$$
 as the main effects  of factors  {\fa}, {\fb}, and {\fc} under weighting vectors
 $\pbc$,  $\pac$, and $\pab$, respectively;  define
$$
\tau_\ab(\pic)  = \sum_{c=0,1} \pi_c \cdot\tau_{\ab|c}, \quad
\tau_\ac(\pib) = \sum_{b=0,1} \pi_b \cdot\tau_{\ac|b},\quad
\tau_\bc(\pia) = \sum_{a=0,1} \pi_a \cdot\tau_{\bc|a}
$$
 as the two-way interactions between factors {\fa} and {\fb}, factors {\fa} and {\fc}, and  factors {\fb} and {\fc} under weighting vectors $\pic $, $\pib $, and $\pia$, respectively; define
$$
\tau_{\abc} =  \tau_{\ab|\co }- \tau_{\ab|\cz } = \tau_{\ac|\bo }- \tau_{\acbz} = \tau_\bcao- \tau_\bcaz 
= \sum_{a,b,c}(-1)^{a+b+c+1} \bY(abc)
$$ 
as the three-way interaction between factors {\fa}, {\fb}, and {\fc}.
\end{definition}

Definition \ref{def_23} gives a total of $2^3-1 = 7$ general  factorial effects, vectorized as
$
\tpi 
= (\tauA(\pbc), \tauB(\pac), \tau_\C(\pab), \tau_{\ab}(\pic), \tau_{\ac}(\pib), \tau_{\bc}(\pia), \tau_{\abc})^\T  = G_\pi \bY.
$
Following the convention from the $2^2$ experiment, the subscripts of the weighting vectors  indicate the factors that are being marginalized out. 
Refer to $\pi$ as the equal weighting scheme if $\pi_{ab}=\pi_{bc}=\pi_{ac}=1/4$ and $\pi_a=\pi_b=\pi_c = 1/2$ for all $a,b,c = 0,1$; 
refer to $\pi$ as the empirical weighting scheme if $\pi_{a} = N^{-1}\sumi \mathcal{I}(A_i = a)$, $\pi_{ab} = N^{-1}\sumi \mathcal{I}(A_i = a, B_i = b)$, etc., equaling the empirical treatment proportions in the experiment.  
Although Definition \ref{def_23} can be general, we focus on the following coherent weighting scheme throughout the paper.

\begin{definition}\label{def:coherent}
A weighting scheme $\pi$ is coherent if there exists a probability distribution over $\mT$,  represented by
$ \pi_{abc} = \pr(A= a, B=b, C=c)$ for $a,b,c=0,1$, 
such that 
\begina\label{te_23_pi}
\begin{array}{lll}
\pi_a=\pr(A = a), &\pi_b =\pr(B = b), &\pi_c =\pr(C = c), \\
 \pi_{ab} =  \pr(A=a,B=b) , \quad & \pi_{ac} =  \pr(A=a, C= c) ,\quad& \pi_{bc} =  \pr(B=b, C= c).
\end{array}
\enda
\end{definition}
 
Coherence imposes mild restrictions on the elements in $\pi$ and, building on the intuition from Remark \ref{rmk:bb}, provides the causal interpretation of the general factorial effects from a thought experiment perspective. 
Consider a target thought experiment in which we assign unit $i$ to combination $(abc) \in \mt$ with  probability $\pr\{Z_i = (abc)\} = \pr(A_i = a, B_i=b, C_i=c) = \pi_{abc}$. 
The weighting vector $\pbc = (\pi_{bc})_{b,c = 0,1}$ gives the marginal distribution of $(B_i, C_i)$  and renders the weighted average $\tau_{\A,i}(\pbc) = \sum_{b,c} \pbca \cdot \tau_{\A| bc, i}$, where $\tau_{\A| bc, i} = Y_i(1bc) - Y_i(0bc)$, an intuitive summary of the main effect of factor A on unit $i$,  accounting for the target treatment probabilities of factors B and C  \cite[see also][]{Hainmueller2014, imai, cuesta2019improving}.  
Averaging $\tau_{\A,i}(\pbc)$ over $i = \ot{N}$ yields $N^{-1}\sumi \tau_{\A,i}(\pbc) = \tau_\A(\pbc)$ as the average effect at the population level. 
The general weights as such allow for  
external validity beyond the actual experiment being conducted. 
The equal weighting scheme  is  coherent with $\pi_{abc} = 1/8$, implying balanced design in the thought experiment. 
The empirical weighting scheme  is also coherent with $\pi_{abc} = e_{abc} = N_{abc}/N$.

\subsection{Factor-based regression with the saturated model}

Define $ A_i'  = A_i - \da$, $B_i'  =B_i - \db$, and $C_i'  = C_i -\dc$ for  prespecified $\bd=(\da,\db,\dc)$ with $ 0\leq \delta_\A, \delta_\B, \dc \leq 1$, and extend \eqref{ols_22} to the $2^3$ experiment to define
\begin{eqnarray}\label{ols_23_g}
Y_i \sim  1 + A^\prime _i + B^\prime _i + C^\prime _i +A^\prime _iB^\prime _i + A^\prime _iC^\prime _i + B^\prime _iC^\prime _i + A^\prime _iB^\prime _iC^\prime _i . 
\end{eqnarray}
Let $\hgd$ and   $\hsd$ be the coefficient vector and robust covariance of the $2^3-1 =7$ non-intercept terms in \eqref{ols_23_g}, respectively. 
We study in this subsection their design-based properties, illustrating two important characteristics of factor-based regressions with more than two factors. 
First, saturated regressions like \eqref{ols_23_g}  can only recover a subset of the coherent factorial effects with   weighting schemes featuring a product structure in Definition \ref{def::product} below. Second, the absence of the three-way interaction restores the generality of \eqref{ols_23_g} for estimating all coherent factorial effects.

\begin{definition}\label{def::product}
A coherent weighting scheme $\pi$ is a product weighting scheme if $\pi_{abc} = \pi_a \pi_b \pi_c$ for $a,b,c=0,1$. 
\end{definition}

 A product weighting scheme $\pi$ is fully determined by the values of  $(\pi_\ao, \pi_\bo, \pi_\co)$ and  implies  independent factors in the corresponding thought experiment. The equal weighting scheme satisfies Definition \ref{def::product} with $\piao=\pibo=\pico=1/2$; the empirical weighting scheme, on the other hand, in general does not.

Let $\dpr$ be the product weighting scheme 
with  $\prh(A_i = 1) = \da$, $ \prh( B_i=1) = \db$, and $ \prh(C_i=1)=\dc$ in the corresponding thought experiment.
As a convention, we use ``$\times$" in  the subscript to indicate product weighting schemes.
Let $\tau_\dprs  = G_\dprs \bY$ be the corresponding vector of general  factorial effects, 
  $\htd = G_\dprs  \hY$ be  its moment estimator, and $\hchtd =  G_\dprs  \hV G_\dprs ^\T$ be the estimated covariance of $\htd$, respectively.
Proposition \ref{prop:ols_23} below gives the numeric correspondence between $\{\hgd, \hsd\}$ and $\{\htau_\dprs, \hchtd\}$, elucidating the utility of \eqref{ols_23_g} for inferring $\tau_\dprs$.

\begin{proposition}\label{prop:ols_23}
{\pretth} 
$
\hgd = \htd 
$  
and
$\hsd  =  \hchtd - G_\dprs \diag(  N_z^{-1})   \hV   G_\dprs^\T.
$ 
\end{proposition}

Proposition \ref{prop:ols_23} highlights the commonality and difference between the $2^2$ and $2^3$ experiments. 
On the one hand,  it ensures the asymptotic equivalence between $\{\hgd, \hsd\}$ and $\{\htau_\dprs, \hchtd\}$ as $N$ goes to infinity, and thereby allows for the large-sample Wald-type inference of $\tau_\dprs$ based on \eqref{ols_23_g}. 
On the other hand, the product structure of $\dpr$ 
constrains the generality of \eqref{ols_23_g}, and suggests that it recovers the full vector of $\tpi$ simultaneously if and only if $(\da,\db,\dc) = (\pi_{\ao}, \pi_{\bo}, \pi_{\co})$ and  $\pi$ is a product weighting scheme. 
The  standard effects satisfy the product structure with $\piao=\pibo=\pico=1/2$  and thus admit of direct estimation with $\da=\db=\dc = 1/2$. 
 
The resulting specification is equivalent to that under the $\{+1,-1\}$ coding system up to a constant scaling factor on each regressor, 
suggesting the   specificity of the $\{+1,-1\}$ coding system  to the standard effects \citep{wh, FactLu16b}.
The partial effects, on the contrary, may or may not satisfy the product structure, and are thus not necessarily directly estimable from \eqref{ols_23_g}; see Remark C1 in the Supplementary Material. This affords a useful guideline for designing and analyzing factorial experiments.

One exception, however, is when the three-way interaction does not exist.
The absence of  $\tau_{\abc}$ leaves the class of  product weighting schemes equivalent to the class of coherent weighting schemes in defining the general factorial effects. We formalize the idea in Proposition \ref{prop:universality_23} below. 
For an arbitrary weighting scheme $\pi$, let $\pi_\prods$ be the product weighting scheme 
with  $\prh(A_i = 1) = \piao$, $ \prh( B_i=1) = \pibo$, and $ \prh(C_i=1)=\pico$ in the corresponding thought experiment. 
By definition, $\pi_\prods$ and $\pi$ share the same marginal treatment probabilities in the underlying thought experiments, and satisfy $\pi_\prods = \pi$ if $\pi$ is already a product weighting scheme. 

\begin{proposition} \label{prop:universality_23}
Under the $2^3$ experiment, if $\tau_{\abc} = 0$, then $\tpi =  \tau_{\pi\prods}$
for all coherent $\pi$, where $\tpi = G_\pi  \bY$ and $\tau_{\pi\prods}  = G_{\pi\prods}\bY$ are the vectors of general factorial effects under weighting schemes $\pi$ and $\pi_\prods$, respectively. 
\end{proposition}

Propositions \ref{prop:ols_23} and \ref{prop:universality_23} together justify the inference of $\tau_\pi$ from    \eqref{ols_23_g} with $(\da,\db,\dc) = (\pi_{\ao}, \pi_{\bo}, \pi_{\co})$ for all coherent $\pi$ when $\tau_\abc=0$. 
The absence of the three-way interaction restores the generality  of  factor-based   regressions for all coherent weighting schemes.

\subsection{Factor-based regression with an unsaturated model}\label{sec::us-23}
Consider  an extension to \eqref{us_22}, 
\beginy\label{us_23}
Y_i \sim 1+  A^\prime _i +  B^\prime _i +  C^\prime _i +A^\prime _iB^\prime _i + A^\prime _iC^\prime _i + B^\prime _iC^\prime _i ,
\endy
when   only  the main effects  and two-way interactions are of interest, vectorized as 
$$
\tpp  =  (\tauA(\pibc), \tauB(\piac), \tau_\C(\piab), \tauAB(\pic),  \tau_\ac(\pib),  \tau_\bc(\pia)  )^\T = \tau_\pi \backslash \{\tau_\abc\} .
$$
Let $\tgd$ and $\tsp $ be the coefficient vector and robust covariance of the six non-intercept terms from \eqref{us_23}.  
We use ``$\tilde{\color{white}{a}}$" to signify outputs from unsaturated regressions, and ``+" to signify quantities associated with the effects of interest throughout the paper.
Let  $\hgp $ and $\hg_{\abc}$ be the coefficients of  $(A^\prime _i ,B^\prime _i , C^\prime _i ,A^\prime _iB^\prime _i , A^\prime _iC^\prime _i , B^\prime _iC^\prime _i)$ and $A_i' B_i' C_i' $ from \eqref{ols_23_g}, respectively, with  $\hgd = (\hgp ^\T, \hg_{\abc})^\T$. Proposition \ref{prop:cf_23} below extends the result in Proposition \ref{prop:cf_22} to the $2^3$ experiment, elucidating the design-based properties of $\tgd$ via its link with $\hg$.

\begin{proposition}\label{prop:cf_23}
Under the $2^3$ experiment, 
we have 
$
\tgd =\hgp   +D   \hg_{\abc}
$
with
\begina
\renewcommand{\arraystretch}{1.2}
D = \left (\sumz e_z^{-1}\right)^{-1} \left( 
\begin{array}{ccc|ccc}
 & &\color{white}1&
 \db & \dc &0 \\
\color{white}1&I_3&&\da & 0& \dc \\
&& &0& \da & \db 
\\\hline
&&&  \\
&0_{3\times3}&&&  I_3\\
&&&&& 
\end{array}\right) 
\left(
\begin{array}{r}
-\sum_{a}e_{a00}^{-1}\\
-\sum_{b}e_{0b0}^{-1}\\
-\sum_{c}e_{00c}^{-1}\\\hline
\sum_{ab} e_{ab0}^{-1}\\
\sum_{ac} e_{a0c}^{-1}\\
 \sum_{bc} e_{0bc}^{-1}
 \end{array}
 \right) - 
 \left (
 \begin{array}{c}
 \db\dc\\
 \da\dc\\
 \da\db\\\hline
  \dc\\
  \db\\
  \da
  \end{array}
  \right).
\enda
\end{proposition}

\def\reg{regression }
\def\regs{regressions }

Recall that $\hgdp$ and $\hg_{\abc}$ equal  the moment estimators of $\twp$ and $\tau_\abc$, respectively, denoted by $\htwp$ and $\htau_\abc$. 
The coefficients from \eqref{us_23} thus recover the exact moment estimator $\htwp$ if and only if $D = 0_6$ or $\htau_\abc=0$. 
The former in general entails $e_z = 1/8$ for all $z\in\mt$  and $\da = \db = \dc = 1/2$, implying both balanced design and standard effects as the estimands.  In particular, $e_z = 1/8 \ (z\in\mt)$ and $\da=\db=\dc=1/2$ ensure that the columns of the design matrix of the saturated regression \eqref{ols_23_g} are mutually orthogonal such that deletion of any subset of the columns has no effect on the estimation of the remaining coefficients, with \eqref{us_23} being a special case. This is in line with the intuition from the $2^2$ case and echos the classical principle that advocates the use of balanced designs whenever possible. 

On the other hand, 
Proposition \ref{prop:cf_23} implies $E(\tgd) -  \twp = D  \tau_\abc$ such that 
$\tgd$ is unbiased for $\twp $ as long as the nuisance effect $\tau_\abc$ indeed  does not exist. 
This, together with the equivalence between $\tpi$ and $\tau_{\pi\prods}$ in the absence of $\tau_\abc$, ensures the  generality of \eqref{us_23} for estimating $\tpp$. 
More precisely, under $\tau_{\abc} = 0$, the coefficient $\tgd$ from \eqref{us_23} with $(\da,\db,\dc) = (\pi_{\ao}, \pi_{\bo}, \pi_{\co})$ is unbiased for $\tpp$ for all coherent weighting schemes $\pi$. 

The violation of the no three-way interaction condition, on the other hand, subjects $\tgd$ to non-diminishing bias, namely $D\tau_\abc$. The intuition on the bias-variance trade-off from the $2^2$ case extends here and ensures that 
$\tgd$ is more precise than $\hgdp$ under Condition \ref{cond:sa} regardless of whether $\tau_{\abc}=0$ or not.

%
\section{A general theory for the $2^K$ factorial experiment}\label{sec:general}
%
\subsection{ Overview and notation}
The $2^K$ factorial experiment features $Q = 2^K$ treatment combinations arising from $K$ binary factors, indexed by $k = \ot{K}$. 
Of interest is the utility of the corresponding factor-based regressions for inferring the factorial effects of interest from the design-based perspective. 
To this end, we first extend the definitions of general factorial effects, coherent weighting scheme, and product weighting scheme to the $2^K$ design, and demonstrate the utility of location-shifted regressions for recovering general effects under product weighting schemes. 
We then show the equivalence between the coherent and product weighting schemes under the no three-way interactions condition.
We finally quantify the bias-variance trade-off between the saturated and unsaturated specifications.

We use the following notation to facilitate the discussion.
Let $Z_{ik} \in \{0,1\}$ denote the level of factor $k$ received by unit $i$ for $ \ i = \ot{N}$ and $k = \ot{K}$. 
Let $\mf_k = \{0, 1\} = \{0_k , 1_k\}$ be the set of possible levels of factor $k$, where we use the subscript $k$ to differentiate the factors.
Let  $  \mt = \prod_{k=1}^K \mf_k =\{z=(z_1, \ldots, z_K): z_k \in\mf_k; \ k=1,\ldots, K\}$ be the set of the $2^K$ treatment combinations.
Let $\mps = \{\mk: \emptyset \neq \mk \subseteq [K]\}$ be the set of all non-empty subsets of $[K]$.
For  $\mk \in\mps$, let $\zk= (z_k)_\kik  $ and $\zkc = (z_k)_{k\not\in\mk}$ index the combinations of factors in $\mk$ and $ \mkc = [K] \backslash \mk$,
 respectively, taking values from $\mfk  = \prod_\kik   \mf_k  = \{0,1\}^{|\mk|}$ and 
$\mf_\mkcs = \prod_{k\not\in\mks} \mf_k = \{0,1\}^{  K - |\mk|}$. 
In particular, $z_\otk = z \in \mt$ and $\mf_{\otk} = \mt$. 

\subsection{Definition of the conditional factorial effects}\label{sec::def_ctau_2k}
Consider $K$ types of factorial effects, quantifying the main effect of a factor when applied alone and the two- to $K$-way interactions when multiple  factors are applied together. 
Refer to them interchangeably as the first- to $K$th-order factorial effects, respectively.  Building on the intuition from the $2^2$ and $2^3$ experiments, we first define the conditional  factorial effects in  this subsection, and  then define the general  factorial effects as their respective weighted averages in the next subsection.

 As a general rule, we define by induction the $m$th-order conditional factorial effect as the difference between two $(m-1)$th-order conditional effects for $m  =  2, \ldots, K$ \citep{wh}. 
For notational simplicity, we illustrate the definition of the $m$th-order effects using the first $m$ factors with $\mk  = [m]$, 
$z_{\mpotk}=  (z_k)_{k=\mm+1}^{K}$, and $\mf_{\mpotk} = \prod_{k=\mm+1}^K \mf_k = \{0,1\}^{K-m}$. 

\begin{definition}\label{def:ct_2k}
Let $\bY(z_1,z_\ttk)$ be the average potential outcome under 
$z = (z_1,z_\ttk) \in \mt$, and define 
$
\tau_1(z_\ttk) = \bY( 1_1 ,  z_\ttk) -  \bY(0_1, z_\ttk)
$
as the conditional main effect of factor 1 when factors $2$ to $K$ are fixed at  $z_\ttk\in \mf_{\ttk}$. 

Given $
\tau_\otmmo ( z_{\ktk})
$ as  the conditional  $(\mm-1)$th-order factorial effect of factors 1 to $(\mm-1)$  when the rest of the factors are fixed at $z_\ktk \in \mf_\ktk$ for $m = 2, \ldots, K-1$, define  
\begina
\tau_\otm (\zotokc)
= \tau_\otmmo (1_m,  \zotokc) -\tau_\otmmo ( 0_m,   \zotokc)  
\enda
as the conditional $\mm$th-order factorial effect of factors 1 to $\mm$ when the rest of the factors are fixed at $\zotokc \in \mf_\mpotk$. 

For $m=K$, define $ \tau _\otk =  \tau_{\otkmo} (1_K) -\tau_{\otkmo} (0_K)   $ as the $K$-way interaction of factors 1 to $K$.
\end{definition}

Based on Definition \ref{def:ct_2k}, we can obtain the  explicit form of $\tau_\otm(\zotokc)$ in terms of the $\by(z)$'s, and show that the order in which new factors are added to the combination in the induction does not matter.  Definition \ref{def:ct_2k} extends to general $\mk \in \mps$ by symmetry.
Denote by 
$\tokzkc$ the conditional $|\mk|$-th order factorial effect of factors in $\mk$ when the rest of the factors are fixed at $\zkc\in \mf_\mkcs$. 
This gives a total of $ |\mf_\mkcs| = 2^{K- |\mk|}$  conditional factorial effects for the $|\mk|$ factors in a fixed $\mk$.
The notation from the $2^2$ case is a special case with
$\tau_{\A | b} =   \tau_\A( b )$ and $\tau_{\B | a} =   \tau_\B ( a )$; likewise for $\tau_{\A | bc} =   \tau_\A ( bc)$, $\tau_{\AB | c} =   \tau_\AB ( c)$, etc. from the $2^3$ case.

\subsection{Definition of the general  factorial effects}\label{sec::def_tau_2k}
We  next  define the general  factorial effects as weighted averages of their respective conditional counterparts. Consider $\pi(z)= \pr(Z_{i1}=z_1, \ldots, Z_{iK}=z_K)$ for $z = (z_1, \ldots, z_K)\in\mt$ as the treatment probabilities under some target thought experiment. 
The marginal distribution of $Z_{i,\mks} = (Z_{ik})_{\kik}$ equals $\pi_\mks = \{\pi(\zk): \zk \in \mfk \}$
with 
$\pi(\zk) = \pr(Z_{i,\mks} = z_\mks )= \sum_{\zkcs \in \mf_\mkcss} \pi(  \zk,  \zkc)$.
It induces an intuitive weighting scheme for averaging over factors in $\mk$ when defining the general factorial effect of factors in $\mkc $.
The $\pia = (\piaz, \piao)$ and $\piab = (\pi_{ab})_{a,b=0,1}$ from the $2^2$ and $2^3$ experiments are both special cases of $\pi_\mks$ with $\mk = \{\text{A}\}$ and $ \{\text{A}, \text{B}\}$, respectively.
Building on the intuition from Definition \ref{def:coherent}, we call $\pi = \{\pi_\mks: \mk \in \mps\}$ the  coherent weighting scheme induced by the joint distribution $\{\pi(z): z\in\mt\}$.

\begin{definition}\label{def:mf_g_2k}
Given a coherent weighting scheme $\pi$ and conditional factorial effects $\tokzkc$ from Definition \ref{def:ct_2k} for all  $ \mathcal{K} \in \mps$ and $\zkc \in  \mf_\mkcs$, define  $ \tkp  =  \sum_{\zkcs\in \mf_\mkcss}\pi(\zkc  )\cdot\tokzkc $
as the general  factorial effect of factors in $\mk$ under $\pi$, vectorized as $
\tpi = \{\tkp : \mk \in\mps \}  = G_\pi \bY$. 
\end{definition}

Definitions \ref{def:ct_2k} and \ref{def:mf_g_2k} together 
define the $(2^K-1)$ general  factorial effects under the coherent weighting scheme $\pi$. 
Refer to $\tkp $ as the standard effect if $\pi(\zkc)= |\mf_\mkcs|^{-1} = 2^{-|\mkcs|}$ are identical for all $\zkc\in\mf_\mkcs$. 
Refer to $\tkp$ as the empirical effect if $\pi(\zkc) = N^{-1}\sumi \mathcal I(Z_{i,\mkcs} = \zkc)$ equals the empirical proportion in the actual experiment.

%
\subsection{Factor-based regression with the  saturated model}\label{sec:saturated model}

Motivated by \eqref{ols_22} for the $2^2$ experiment and \eqref{ols_23_g} for the $2^3$ experiment, we define $Z'_{ik} = Z_{ik} - \delta_k$ and $Z'_{i, \mks} = \prod_{\kik } Z'_{ik}$ as a location-shifted generalization for some prespecified $ (\delta_k)_{k=1}^K$ with $0\leq \delta_k \leq 1$, and consider the saturated factor-based regression
\beginy\label{ols_2k_g}
Y_i 
\sim  1 + \sum_{k=1}^K Z'_{ik} + \sum_{1\leq k\neq k'\leq K} Z'_{ik}Z'_{ik'}+ \dots + 
\prod_{k=1}^K Z'_{ik}
 \sim 1+ \sum_{\mk \in \mps} Z'_{i, \mks} .
\endy
Let $\hgd$ and $\hsd $ be the coefficient vector and robust covariance of the $(Q-1)$ non-intercept terms in \eqref{ols_2k_g}, respectively, with elements arranged in the same order of $\mk$'s as those in $\tpi$. 
We derive below their utility for the Wald-type inference of $\tau_\pi$.

To begin with, the notion of product weighting scheme extends naturally to the current setting as
$
\pi(z) =\prod_{k=1}^K   \pi(z_k)  , 
$
and is fully determined by the values of $\{\pi(1_k)\}_{k=1}^K$. 
The equal weighting scheme for the standard effects satisfies the product structure with $\pi(1_k) = 1/2$. The empirical weighting scheme, on the other hand, may not.  
Building on the intuition from the $2^3$ experiment, Definition \ref{def:prod_2k} below introduces two product weighting schemes of particular importance, arising from the estimand of interest and the location-shift parameters, respectively. 

\begin{definition}\label{def:prod_2k}
For an arbitrary coherent weighting scheme $\pi$, let $\pi_\prods$ 
be the product weighting scheme with $\pi_\prods(1_k) = \pi(1_k)$ for $k = \ot{K}$. 

For arbitrary location-shift parameters $(\delta_k)_{k=1}^K$ with $0\leq \delta_k \leq 1$, let $\dpr$ be the product weighting scheme with $\dpr(1_k) = \dk$ for $k = \ot{K}$. 
\end{definition}

The product weighting scheme $\pi_\prods$ satisfies $\pi_\prods = \pi$ if $\pi$ is already a product weighting scheme. 
The product weighting scheme $\dpr$ features 
$
\dpr(z) = \prod_{k=1}^K\delta_k^{z_k} (1-\delta_k) ^{1-z_k} 
$
for all   $z \in\mt$.
Let $\tau_\dprs  = G_\dprs \bY$ be the corresponding vector of general  factorial effects,  $\htd= G_\dprs \hY$ be its moment estimator, and $\hchtd = G_\dprs  \hV G^\T_\dprs $ be the estimated covariance, respectively.
Theorem \ref{thm:2k_g} below gives the numeric correspondence between $\{\hgd, \hsd\}$ and $\{\htd, \hchtd\}$ for inferring $\tau_\dprs $. 

\begin{theorem}\label{thm:2k_g}
{\pretk}   
$\hgd = \htd$ and $\hsd  =  \hchtd   - G_\dprs \diag(  N_z^{-1})   \hV   G_\dprs^\T$. 
\end{theorem}

Theorem \ref{thm:2k_g} unifies the results from the $2^2$ and $2^3$ experiments and justifies the utility of $\hgd$ and $\hsd$ from \eqref{ols_2k_g} for inferring $\tau_\pi$ when $\pi$ is a  product weighting scheme  with $\pi(1_k)=\delta_k$ for $k = \ot{K}$.  Despite the constrained applicability in general, the intuition from Proposition \ref{prop:universality_23} extends here and ensures the generality of \eqref{ols_2k_g} in the absence of three-way interactions.

\begin{condition}
\label{cond:no3_2k}
Assume  $\tokzkc = 0$ for all $\zkc$ with $|\mk| = 3$. 
\end{condition} 

Condition \ref{cond:no3_2k} rules out the existence of three-way interactions and thus that of all $m$-way interactions for $3 < m \leq K$ by Definition \ref{def:ct_2k}. 
We will refer to it as the no three-way interactions condition hence for simplicity.  

\begin{theorem} \label{thm:universality_2k}
Under the $2^K$ experiment and Condition \ref{cond:no3_2k},  we have $\tpi = \tau_{\pi\prods}$ for all coherent $\pi$,
where $\tpi $ and $\tau_{\pi\prods}$ are the vectors of general  factorial effects under $\pi$ and $\pi_\prods$, respectively. 
\end{theorem}

Theorems \ref{thm:2k_g} and \ref{thm:universality_2k} together allow us to use \eqref{ols_2k_g} with $\dk=\pi(1_k)$ for the Wald-type inference of all $\tpi$ with coherent $\pi$ in the absence of three-way interactions. 
The proof of Theorem \ref{thm:universality_2k} further shows that the requirement of $\tokzkc=0$ for all $|\mk|=3$ is not only sufficient but also  necessary for $\tau_\pi =   \tau_{\pi\prods}$ to hold if $\pi$  is coherent but not a product weighting scheme.
Thus, we cannot relax the $|\mk|=3$ in Condition \ref{cond:no3_2k} to $|\mk|=m$ for some $m>3$ for Theorem \ref{thm:universality_2k} to hold.

%
\subsection{Factor-based regression with unsaturated models}\label{sec:unsaturated model_2k}

Motivated by \eqref{us_22} for the  $2^2$  experiment and \eqref{us_23} for the $2^3$ experiment, 
we next consider 
\beginy\label{us_2k}
Y_i \sim 1+\sum_{\mk \in \mfp} \zpki,
\endy
where $\mfp \subset \mps$, 
as an unsaturated variant of \eqref{ols_2k_g} when only a subset of the $(Q-1)$ factorial effects are of interest, vectorized as $\tpp  = \{\tkp  : \mk \in \mf_\ps   \}$.
A commonly-used special case is
$
Y_i 
\sim 1 +   Z_{i1}' + \dots + Z_{iK}' 
$
with only the first-order terms and $\mfp = \{\{k\}: k = \ot{K}\}$. 
The additive form ensures that the location-shift transformation has no effect on the estimation of the non-intercept coefficients. Another commonly-used special case is
$
Y_i 
\sim  1 + \sum_{k=1}^K Z'_{ik} + \sum_{1\leq k\neq k'\leq K} Z'_{ik}Z'_{ik'}
$
with only the main effects and two-way interactions and $\mfp = \{\{k\}, \{k,k'\}:   k,  k' = \ot{ K} \text{ with }\ k \neq k'\}$. 

Let $\tgp$ and $\tsp $ be the coefficient vector and robust covariance  of the $|\mf_\ps|$ non-intercept terms in  \eqref{us_2k}, respectively. 
We establish in this subsection their utility for inferring $\tpp$. 
Recall $\hgd$ as the coefficient vector of the non-intercept terms from \eqref{ols_2k_g}.  Partition it into $ \hgdp$ and $  \hgdm$, corresponding to the coefficients of $(\zpki)_{\mk \in \mfp}$ and $(\zpki)_{\mk \not\in \mfp}$, respectively.
As a convention, we use ``$+$" and ``$-$" in the subscripts to signify effects included in and omitted from the unsaturated regression \eqref{us_2k}, respectively.

Let $F$ be the $N\times Q$ design matrix of \eqref{ols_2k_g}, concatenating columns of $1_N$ and $(\zpki)_{i=1}^N$ for all $\mk\in \mps$.  
Let $\fp $ be the $N \times (1+|\mfp|)$ design matrix of \eqref{us_2k},
and $ \fm =  F \backslash \fp$ be the submatrix of $F$ omitted from \eqref{us_2k}, concatenating columns of $(\zpki)_{i=1}^N$ for $\mk\not\in \mfp$.   Assume throughout that the elements in $\tgp$, $\hgdp $, and $\hgdm $ are  arranged in the same relative order of $\mk$ as those in $\tpi$; likewise for the columns in $\fp $ and $F_\ms$.

Let $\fmh = {(\fpt \fp)^{-1} \fpt \fm}$ be the coefficient matrix from the column-wise regression of $\fm$ on $\fp$, which is 
a deterministic function of $(e_z)_{z\in \mt}$ by Lemma B4 in the Supplementary Material. 
Let $R = \fm - \fp \fmh$ 
be the corresponding residual matrix, $D$ be the submatrix of $\fmh$ without the first row, 
and $\fpp$ be the submatrix of $\fp$ without the first column.
Let $  P_N $ be the projection matrix orthogonal to $1_N$, and $Y = (Y_1, \ldots, Y_N)^\T$ be the vector of the observed outcomes.
Theorem \ref{thm:cf_2k} below states the numeric correspondence between $\tgd$ and $\hgd$ under the $2^K$ factorial experiment,
generalizing Propositions \ref{prop:cf_22} and \ref{prop:cf_23}.

\begin{theorem}\label{thm:cf_2k}
Under the $2^K$ experiment,  the coefficients from \eqref{ols_2k_g} and \eqref{us_2k} satisfy 
$
\tgp = \hgdp + D   \hgdm 
$, where $ D   \hgdm=0$ if and only if $  \fppt   P_N  \fm (R^\T R)^{-1}R^\T Y= 0 $. 
Two sufficient conditions for $ D   \hgdm=0$ for all $Y$ are  $ \fpt \fm=  0$ or  $\fp^\T P_N \fm = 0$. 
\end{theorem}
 
\def\tkd{\tau_{\mks,\dprs}}
Recall that $\hgdp$ and $\hgm$ coincide with the  moment estimators of $\twp   = \{\tkd: \mk \in \mf_\ps\}$ and $\tau_{\dprs,\ms} =\{\tkd:\mk\not\in\mfp\}$, respectively, denoted by $\htwp$ and $\htau_{\dprs,\ms}$. 
Theorem \ref{thm:cf_2k} gives two sufficient conditions for $\tgp$ to recover exactly $\htwp$, requiring orthogonality of $\fp$ and $\fm$ either in original form or after  centered by the column averages. These conditions do not hold in general unless the design is balanced and the factorial effects are the standard ones under the equal weighting scheme.
This generalizes the intuition from the $2^2$ and $2^3$ cases to general $K$. 

\begin{corollary}
\label{corollary::orthogonality_2k} 
Under the $2^K$ experiment,  
$\tgp = \hgdp$ if \textup{(i)} $\delta_{k}= 1/2$ for all $k = \ot{K}$ and \textup{(ii)} $N_z = N/Q$ for all $z \in \mt$. 
\end{corollary} 

We can drop the balance condition (ii) in Corollary \ref{corollary::orthogonality_2k} if we use the weighted least squares fit with weights $1 / N_{Z_i}$ for $i=1,\ldots, N$. We relegate the details to \S A.5 in the Supplementary Material and focus on the ordinary least squares fit in the main text. 

Despite the loss of exact recovery of the moment estimator $\htwp$ when $D\hgm \neq 0$, the intuition from the $2^2$ and $2^3$ experiments extends here and ensures the unbiasedness of $\tgp$ in the absence of the nuisance effects.

\begin{condition}
\label{cond:noeff_2k}
The nuisance effects are zero, that is, $\tkp  = 0$ for all $\mk \not\in\mfp$. 
\end{condition}

\begin{theorem}\label{thm:nonu_2k}
Under the completely randomized $2^K$ experiment, the  coefficients from \eqref{us_2k} satisfy
$  E(\tgp) = {\twp} + D {\twm}$ and $\cov(\tgp) = ( I,  D) G_\dprs \cov(\hY) G_\dprs^\T ( I,  D)^\T$.
Further assume  Condition \ref{asym} and Condition \ref{cond:noeff_2k} with $\pi=\dpr$.
Then $  E(\tgp) = {\twp}$, and $\tgp$ is asymptotically Normal with 
$
N\{\tsp  - \cov(\tgp)\}=  \Delta + \op 
$, 
where $\Delta =  ( I,  D)   G_\dprs  SG_\dprs ^\T  ( I,  D)^\T\geq 0$. 
\end{theorem}

Theorem \ref{thm:nonu_2k} justifies the Wald-type inference of $\twp$ from the unsaturated specification \eqref{us_2k} when the nuisance effects omitted indeed do not exist. 
The resulting $\tgp$ is both unbiased and consistent for estimating $\twp$, with the robust covariance $\tsp $ affording an asymptotically conservative estimator for the true sampling covariance. 
The proof of Theorem \ref{thm:2k_g}  further shows that the intercept from \eqref{us_2k} is an unbiased estimator of a weighted average of $\bar Y(z)$ instead of a contrast and is thus non-zero in general. 
This suggests the necessity to include the intercept in the unsaturated specification for the satisfaction of Condition \ref{cond:noeff_2k}. 
One limitation of \eqref{us_2k}, again, lies in its requirement on the product weighting scheme. 
Juxtaposing Condition \ref{cond:no3_2k}, Condition \ref{cond:noeff_2k},  and Theorem \ref{thm:universality_2k} together ensures that the result  of Theorem \ref{thm:nonu_2k}  extends to $\tpp$ for all coherent $\pi$ in the absence of three-way interactions.

\begin{remark}\label{remark::tradeoff} The constant treatment effects condition further ensures $
\cov(\tgp)\leq   \cov(\hgdp)
$ such that the estimator from \eqref{us_2k} has smaller sampling covariance compared with that from \eqref{ols_2k_g}. 
This, together with Theorem \ref{thm:nonu_2k}, illustrates the bias-variance trade-off between the saturated and unsaturated  regressions  from the design-based perspective. This result, however, does not hold without the constant treatment effects assumption. We give a counterexample in \S A.4 in the Supplementary Material. 
\end{remark} 

The assumption of no nuisance effects can never be verified exactly in practice.
Extra caution is thus needed when applying unsaturated specifications to unbalanced designs or estimands other than the standard effects.
The saturated specification is, in this sense, a safer choice when the sample size permits.
When the number of treatment combinations $Q = 2^K$ is large relative to the sample size $N$, however, the saturated regression is subject to substantial finite-sample variability, and leaves the unsaturated regressions possibly more attractive alternatives for finite-sample inference.  
Even if the nuisance effects are not exactly zero, depending on our belief of the data generating process, the gain in finite-sample precision by the unsaturated regressions 
can still outweigh the bias as long as the omitted nuisance effects, most likely some higher-order interactions, are reasonably small, ensuring a smaller mean squared error overall.

Alternatively, lasso and ridge regression
afford attractive alternatives when the saturated regression is not possible. 
Indeed, discussion so far holds with a given unsaturated specification   \eqref{us_2k}. It is desirable to have a data-driven specification with both model selection and  post-selection inference \citep{Chipman1997, espinosa2016bayesian, imai}. 
Although these topics have been discussed extensively under the classic linear model, analogous results are largely unexplored under the design-based framework. We leave this to future work.

\section{Discussion and recommendations}

We wrap up this article with three practical implications of our findings in terms of the $2^K$ experiment. The intuition extends to the general $\qg$ experiment with minimal modification as shown in \S A of the Supplementary Material.  
First, 
the definition of the general factorial effects greatly broadens the range of estimands that could be considered under factorial experiments, enabling flexible weighting schemes to accommodate context-specific concerns. 
Second, location-shifted factor-based regression affords a convenient way to recover the moment estimators of the general factorial effects from least squares, with the corresponding robust covariance being an asymptotically conservative estimator of the true sampling covariance. 
This enables large-sample Wald-type inference from least-squares outputs. 
With more than two factors, factor-based regression is capable of estimating general factorial effects under product weighting schemes, yet regains generality in the absence of three-way interactions.
Third, unsaturated regressions reduce sampling variances under the constant treatment effects assumption, but are subject to non-diminishing biases when the no nuisance effects condition is violated.
Importantly, our theory is design-based without requiring any stochastic models for the potential outcomes.

We focused on complete randomization due to its own wide applications. Clarifying the above 
important issues in this basic experiment affords a proof of concept for other more complex experiments.
The definitions of the general  factorial effects remain unchanged, and the correspondence between the least-squares outputs and  moment estimators is purely numeric and thus holds under any randomization mechanism. 
The appropriateness of the Wald-type inference, on the other hand, is assignment specific and requires modifications under different randomization mechanisms.
We conjecture that the theory extends to experiments with non-constant treatment probabilities \citep{mukerjee2018using} if we weight the least-squares procedure by the inverse of the treatment probability. We leave this to future work.

\section*{Supplementary material}
 The Supplementary Material contains the results for the general $\qg$ factorial experiment, details omitted in the main text, and numerical examples. 

\section*{Acknowledgment}
We thank the three reviewers, Avi Feller, Cheng Gao, and Nicole Pashley for constructive comments. Peng Ding was partially supported by the U.S. National Science Foundation.

\bibliographystyle{plainnat}
\bibliography{refs_PA}

\end{document}